\documentclass[aps,prl,twocolumn, superscriptaddress,showpacs]{revtex4}
\usepackage{graphicx}
\usepackage{psfrag}
\usepackage{subfigure}
\usepackage{verbatim}
\usepackage{color}
\newcommand{\bea}{\begin{eqnarray}}
\newcommand{\eea}{\end{eqnarray}}

\begin{document}


\title{Dielectric breakdown and avalanches at non-equilibrium metal-insulator transitions}


\author{Ashivni Shekhawat}
\author{Stefanos Papanikolaou}
\affiliation{LASSP, Physics Department, Clark Hall, Cornell University, Ithaca, NY 14853-2501}
\author{Stefano Zapperi}
\affiliation{Consiglio Nazionale delle Ricerche-IENI, Via R. Cozzi 53, 20125 Milano, Italy}
\affiliation{ISI Foundation, Viale S. Severo 65, 10133 Torino, Italy}
\author{James P. Sethna}
\affiliation{LASSP, Physics Department, Clark Hall, Cornell University, Ithaca, NY 14853-2501}


\date{\today}

\begin{abstract}
Motivated by recent experiments on the finite temperature
Mott transition in VO$_2$ films, we propose a classical
coarse-grained dielectric breakdown model where each degree of freedom
represents a nanograin which transitions from insulator to metal with
increasing temperature and voltage at random thresholds due to quenched
disorder. We describe the properties of the resulting non-equilibrium
metal-insulator transition and explain the universal characteristics of the
resistance jump distribution. We predict that
by tuning voltage, another critical point is approached,
which separates a phase of ``bolt"-like avalanches from percolation-like
ones.
\end{abstract}




\maketitle

Vanadium dioxide~(VO$_2$), when heated or strained, displays an insulator to metal transition with intriguing non-equilibrium collective behavior,  portrayed in a remarkable series of recent experiments~\cite{sharoni08,chang07,qazilbash09,wu2006,wu2011}. Strong electron correlations drive the microscopics of this metal-insulator transition, where a delicate interplay among structural, electronic and spin degrees of freedom takes place~\cite{imada98}. However, as we argue in this Letter, the universal features of the observed resistance jumps can be understood via appropriate generalizations of previously studied models of dielectric breakdown~\cite{duxbury1986, niemeyer1984}. By tuning two natural control parameters, the applied voltage $V$ and the contrast $h$ (the ratio of conductances of the insulating and metallic domains), we show that the existing experiments are in the small $h$ regime, where a crossover, in small samples, between a low-$V$ percolating phase and a high-$V$ ``bolt" phase takes place. As $h$ becomes larger, this crossover evolves to a sharp transition with novel critical properties.

\begin{figure}[t]
\includegraphics[width=0.4\textwidth,angle=0]{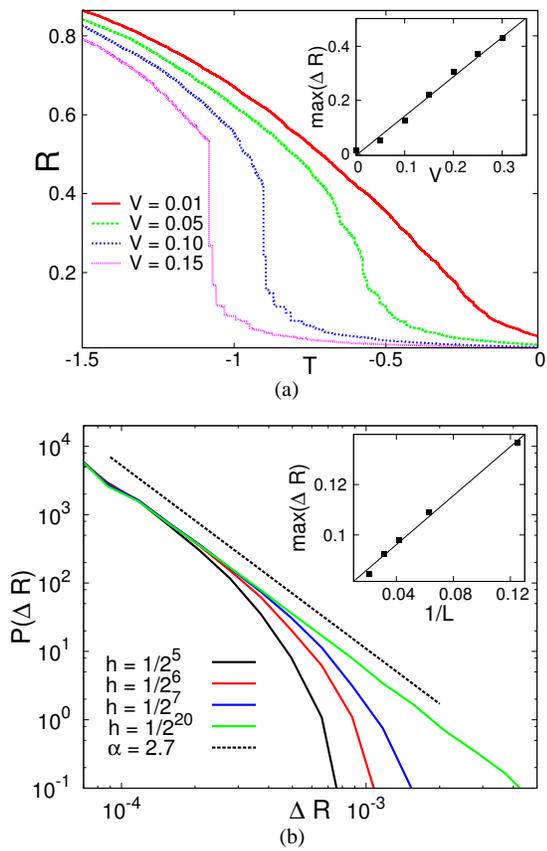}
\caption{{\bf Universal scaling and avalanches in the high-contrast, percolation-dominated regime}. 
(a) The resistance-temperature curve shows a multiple-step structure similar to the experimentally observed one~\cite{sharoni08}. 
(b) The resistance jump 
distribution acquires a universal form, for different contrast 
parameters and voltages for $L=128$. The exponent $\alpha=2.7$ agrees qualitatively with the experimentally observed exponent 2.45.
Additionally, the distributions show finite-size scaling, demonstrating the presence
of a nearby critical point. 
{\bf Insets}: The largest resistance jump is observed to scale linearly with $1/L$ at fixed $V=0.1$~(inset in (b), as observed experimentally~\cite{sharoni08}), and linearly with $V$ at fixed $L=128$~(inset in (a)) .}
\label{fig:HC}
\end{figure}

The VO$_2$ films studied in Ref.~\cite{sharoni08} had a thickness of 
90~nm, widths ranging from 2~$\mu$m to 15~$\mu$m and lengths ranging from 200~nm to 4~$\mu$m.  X-ray diffraction studies of films near criticality revealed that stable insulating grains 
have an average linear size of 20~nm~\cite{chang07,kikuzuki2010}.
With the sample put under an external voltage $V$, multiple resistance jumps were observed 
near the bulk transition temperature~\cite{sharoni08}. The statistics of these jumps revealed a power law probability distribution 
$P(\Delta R)\sim \Delta R^{-\alpha}$, with an exponent $\alpha\simeq2.45$. The resistance jump distribution depended strongly 
on the magnitude of the external voltage, with the largest jump scaling  linearly with the voltage. 
Further, in the presence of external voltage, 
elongated conducting clusters have been observed through X-ray diffraction~\cite{chang07}, 
whereas in the absence of voltage, percolation-like isotropic clusters have been recorded with near-field infrared 
spectroscopy~\cite{qazilbash07, qazilbash09}. 
\par

Even though VO$_2$'s transition properties are dominated by electron correlations, 
we argue that the observed collective phenomena can be explained in a purely classical way, consistent with experimental observations~\cite{sharoni08,chang07,qazilbash09,wu2006,wu2011}. 
The large length scales of the domains ($\sim$15-20~nm) and the small electron mean-free path near the transition ($\sim$0.26~nm)
 suggest that coherence effects are unimportant and electron transport is predominantly classical (Ohm
's law)~\cite{footnote3}. 
This high-temperature transition ($\sim$340~K) cannot be interpreted as a quantum phase transition, since the observed $\sim$1\% lattice distortion suppresses any electronic coherence~\cite{footnote2}.
The thermal loading must be considered quasi-static because the loading rate of the experiments ($<3$ K/min,~\cite{sharoni08}) is much slower 
than the intrinsic dynamics of the domains ($\sim$$10^{-3}$ s,~\cite{kikuzuki2010}).
Also, some experiments at high voltage show a large event that repeats in space~\cite{kim07} and time \cite{wu2006} over repeated cycles of thermal loading, while others~\cite{sharoni08}, for smaller voltages do not exhibit this repetition. 
In our model, we consider a quasi-static model of classical resistors in two dimensions with deterministic dynamics, and with classical, quenched disorder, hence leading to reproducible avalanche sequences. The strongly-correlated quantum and statistical physics underlying the Mott transition is absorbed into 
temperature and voltage dependencies of our domain dynamics, which could be estimated by using DMFT methods~\cite{martin2010,eyert2002}.

Motivated by previous successful studies of strongly-correlated
electronic systems at finite temperatures~\cite{salamon01, dagotto-2001,papanikolaou08}, 
we propose an extended dielectric breakdown network model of coarse-grained regions transforming from 
insulator to conductor
with random critical temperature thresholds.
We study the resistance jump distribution and make predictions about the exponent $\alpha$. 
In addition, we study the probability distribution of avalanche sizes $P(s)\sim s^{-\tau}$, where $s$ is the number of resistors transformed 
in a single avalanche burst.
We explain the observed qualitative behavior
at different voltages, and predict the existence of two distinct regimes: 
a) a percolation dominated regime~\cite{roux87} where scaling
appears only in resistance jumps and avalanches are isotropic and small, and, b) a bolt dominated regime, where avalanches are highly anisotropic, almost line-like. 
Finally, we make a number of experimentally verifiable predictions 
regarding the behavior of the system in the different regimes.

In our model each link $i$ of the network, labelled by a
variable $S_i$, is thought of as a microscopic ``grain" of linear size at least
of the order of the dephasing length $l_\phi$. It can be conducting
($S_i\equiv +1$) with conductance $\sigma_C$, or insulating ($S_i\equiv
-1$) with conductance $\sigma_I=h~\sigma_C$.
The variable $0 \leq h \leq 1$ is the inverse contrast between conducting and insulating
regions. We enforce bi-periodic boundary conditions on a diamond lattice (a square grid 
rotated by $45^\circ$) and subject it to an external voltage $V$ per
link. Experimental observations show that the threshold temperature in the insulator to metal transition decreases with voltage \cite{martin2010,kim07,wu2011}. We account for this in the model by transforming the resistor at link $i$ from insulator to metal when the following condition is satisfied,
\begin{eqnarray}
T  \ge T_i^{c}-b V_i
\label{eq:cond}
\end{eqnarray}
where $T$ is the temperature of the sample, $V_i$ is the voltage drop across the $i^{\mathrm{th}}$ link, and $T_i^c$ is the random 
zero-voltage critical temperature threshold which models the disorder~\cite{footnote5}. Equation~\ref{eq:cond} is a linear approximation 
to the observed voltage dependence of the critical temperature threshold~\cite{martin2010, kim07, wu2011}; 
the exact functional form should be irrelevant for the universal behavior.

%

\begin{figure}[t]
\includegraphics[width=0.4\textwidth,angle=0]{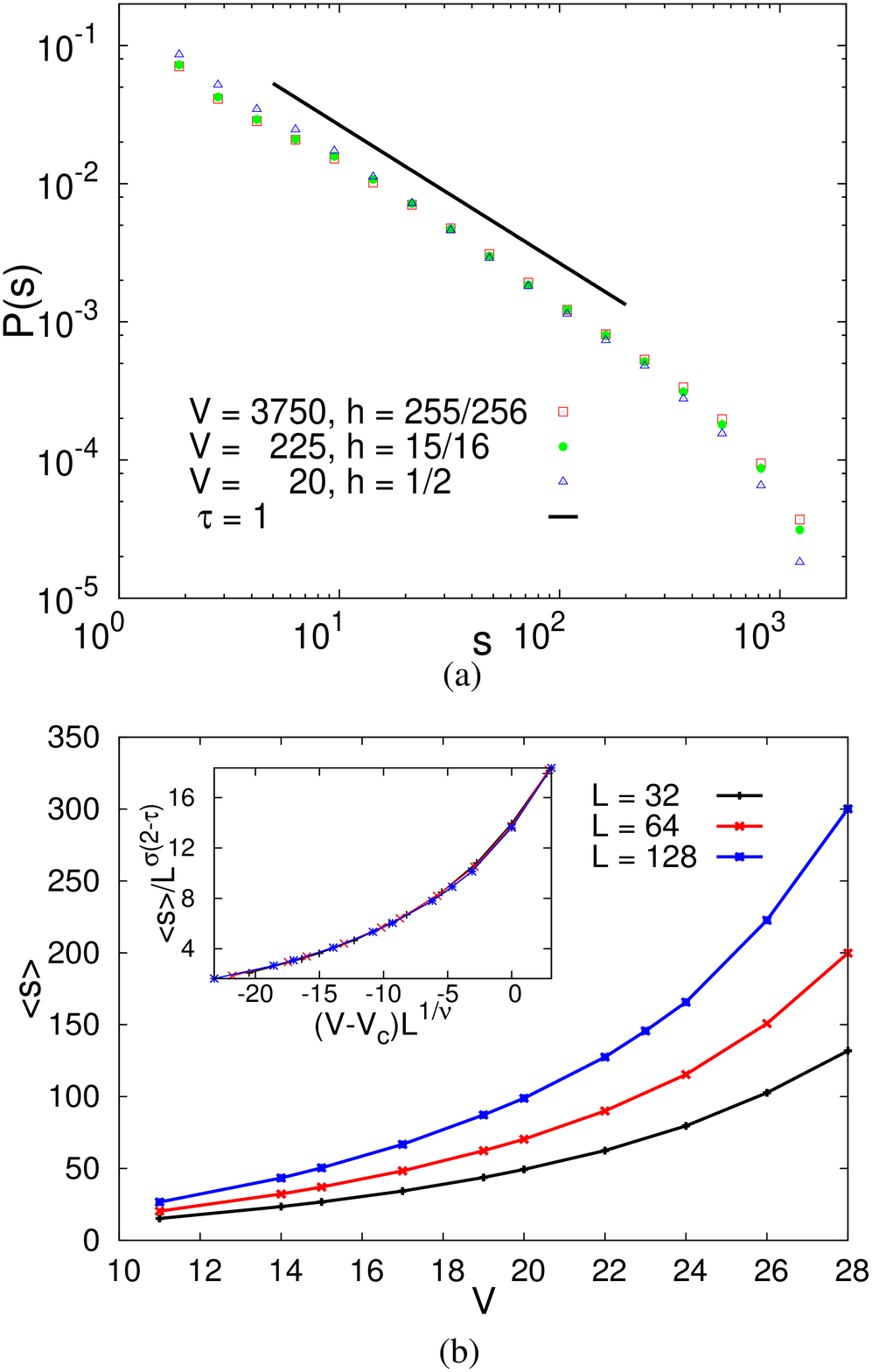}
\caption{{\bf Novel universality at the percolation-bolt transition.} 
(a) The probability distribution of avalanche sizes shows universal scaling near $h = 1$; In this limit, $\Delta R\sim S$.
(b) As shown in the inset, the mean avalanche size at the critical voltage diverges as $L\rightarrow\infty$, suggesting a continuous phase transition. 
}
\label{fig:LC}
\end{figure}

In this model there are two cases which have been studied previously: $V=0$
and the limit $h\rightarrow1$.  
At $V=0$, resistors are not coupled and transform sequentially one at a time
as in percolation. The resistance jump distribution for percolation,
originally studied in Ref.~\cite{roux87}, displays a multifractal structure with
a power law tail at large jumps decreasing with an exponent $\alpha \simeq 2.7$ (the power law tail  is shown in Fig.~\ref{fig:HC}(b)).
As $h\rightarrow 1$, 
the model can be studied by an explicit perturbation expansion in powers of $(1-h)/h$~\cite{blackman76}. The voltage $V_i$ across 
the $i^{\mathrm{th}}$ link satisfies the recursive equation
$V_i = V - \frac{1}{2}[(1-h)/h]\sum_j \Gamma_{ij}(S_j+1)V_j$, where $\Gamma_{ij}$ are the lattice Green functions with dipolar form at long distances.
Their general form for a $n$-dimensional hyper-cubic lattice is  
$\Gamma_{ij} = \int d^n k/(2\pi)^n \sin\frac{1}{2}k_i\sin\frac{1}{2}k_j\cos({\bf k}.{\bf r}_{ij})/(\sum_{l=1}^n\sin^2\frac{1}{2}k_l)$,
where $k_i, k_j$ are the directional wave vector components, and ${\bf r}_{ij}$ is the vector 
from the center of link $i$ to center of link $j$. Taking $\epsilon \equiv (1 - h)/h << 1$ we obtain~\cite{blackman76},
\vspace{-1pt}
\begin{equation}
V_i - V = -(\epsilon V/2){\Sigma}_j\Gamma_{ij}(S_j+1) + \mathcal{O}(\epsilon^2).
\label{eq:perturbation}
\end{equation}
\vspace{-1pt}
Thus, in the singular limit of $h\rightarrow 1$, the model maps to a disordered, long-range, frustrated Ising model. 
This mapping is intriguing, because it maps a dielectric breakdown model with {\it non-additive} multi-body interactions, to a dipolar Ising model with 
additive two-body interactions. The dipolar interaction in this singular limit is shared with a model~\cite{mughal2009} of interface depinning in magnetic hysteresis, where their finger-like structures resemble our bolts.


We perform numerical simulations, where a random temperature threshold $T_i^c$, drawn from the standard Gaussian distribution, is assigned to each link. The simulation starts with every resistor in the insulating state ($S_i = -1\ \forall\ i$). The voltage at individual nodes is found by numerically solving the Kirchoff equations~\cite{nukala2003}. At each step the resistor for which the condition $T  = T_i^c - b|V_i|$ (Eq.~\ref{eq:cond}) is satisfied at the lowest possible value of $T$, is transformed into metal, and voltages are recomputed for the entire network. 
The process is repeated until every resistor in the conducting  state ($S_i = 1\ \forall\ i$). 

\par

In the experiments of Refs.~\cite{sharoni08, chang07, qazilbash07} on VO$_2$, $h$
is small (about $10^{-3}$) and the voltage appears to be low compared to the disorder threshold.
In this limit, for  large resistance jumps, shown in Fig.~\ref{fig:HC}(a), the distribution has an exponent $\alpha\simeq2.7$ which is very
similar to the experimental findings reported in Ref.~\cite{sharoni08}. The structure of the resistance-temperature curve shown in 
Fig.~\ref{fig:HC}(b) is also similar to ones reported experimentally. The size of the largest resistance jump scales linearly with 
the applied voltage, as reported in Ref.~\cite{sharoni08}. 
This dependence on the applied voltage stems from the non-additive multi-body interactions of our model, and cannot be 
achieved by previously suggested bond-percolation type models~\cite{sharoni08} where the size of the 
largest resistance jump vanishes in the large system size limit. 
A more explicit signature of percolation would be the observation of the 
multifractal scaling \cite{roux87} expected at low resistance jumps, possibly below the experimental resolution. 

When the contrast is smaller ($h\gtrsim 1/2$), we find that the insulator to metal transition occurs in avalanches, with several bonds 
transforming simultaneously at the same temperature. For fixed $h$ (near $1$), avalanches and resistance jumps are 
linearly related ($\Delta R \sim s(1-h)/(2 L^2)$ for the diamond lattice)
and both show power laws and universal scaling (sizes shown in Fig.~\ref{fig:LC}).
As the external voltage $V$ is 
varied, the avalanche size distribution evolves from trivial (at $V = 0$, where resistors transform one by one) to a power law at a critical voltage $V_{\rm cr}(h)$, to again trivial (one giant avalanche) at $V \gg V_{\rm cr}$. This behavior is suggestive of a continuous phase transition; we analyze the probability distribution of our sizes $P(s)$ with the
scaling form $P(s) \sim s^{-\tau} \Phi(s/\xi^\sigma,L/\xi)$, where $\xi \sim |V - V_{\rm cr}|^{-\nu}$ is the correlation length.
The $n^{th}$ moment of the avalanche size distribution scales as 
$\langle s^n \rangle \sim L^{\sigma(1+n-\tau)}\Psi((V-V_{\rm cr})L^{1/\nu})$. These scaling forms fit the data with good accuracy as shown in
Fig.~\ref{fig:LC}. Figure~\ref{fig:LC}(a) shows the universal size distribution and Fig.~\ref{fig:LC}(b) shows the distribution of 
the mean avalanche size, and a fit to the predicted scaling form. From these fits we get $1/\nu = 0.25\pm0.24,\ \sigma=0.8\pm0.4,\ \tau = \alpha = 1\pm0.2$. 
We have also studied other disorder distributions (e.g.~$T_i^c$ taken from a uniform or exponential distribution) and explored other 
analytical methods (e.g.~changing the critical range in the fits and analyzing the size distribution of spatially 
connected pieces of the avalanches) all of which confirm the presence of critical fluctuations. 

\begin{figure}[t]
\begin{center}
\includegraphics[width=0.45\textwidth]{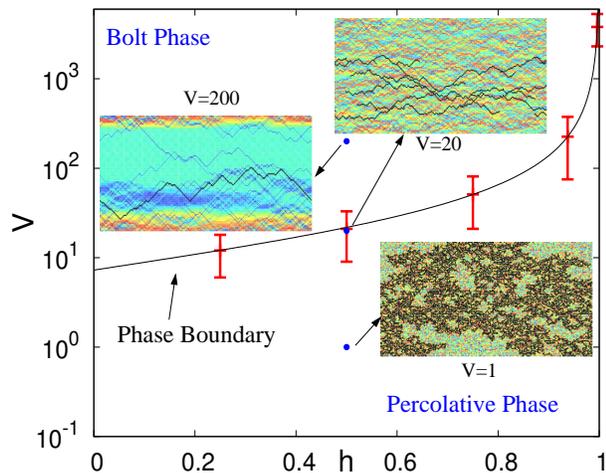}
\caption{{\bf Fractal-looking clusters and phase diagram}. The colors in the insets reflect avalanches; the first spanning cluster is shown in black. 
The phase boundary and the error-bars are obtained by treating the critical voltage, $V_{cr}$, as
a free parameter in data collapses (see Fig.~\ref{fig:LC}).
The percolation fixed point at $h=0, V=0$ is likely unstable under
coarse-graining, and we anticipate that there will be a crossover to
the critical point ($h\rightarrow 1$) behavior for very large avalanches even for high contrast.}
\label{fig:PT}
\end{center}
\end{figure}

The phase transition identified above separates a percolative phase from a `bolt' phase as shown in Fig.~\ref{fig:PT}.
We estimate the phase boundary by a mean-field theory that becomes exact in the limit of $h\to 1,\ V\to 0$. In this limit the local voltage concentration are unimportant and the interactions are additive. The avalanches can be  modeled as a branching process -- a grain (bond) turning metallic induces a long-ranged perturbation in the voltage field, which 
can result in a few more grains turning metallic, ad infinitum. The voltage change, $\Delta V$, due to a single metallic bond  at a distance $r$,  goes as $\Delta V(r) \propto \frac{V(1-h)}{r^2(1+h)}$ \footnote{A continuum result obtained by  approximating a single metallic bond by a circular inclusion in a 2-D domain.}. 
Let $\lambda$ be the average number of grains that turn metallic due the perturbation caused by one grain, then
$\lambda \propto V\log L (1-h)/(1+h)$. The mean size of the resulting avalanche is 
given by $1 + \lambda + \lambda^2 + \ldots$. Thus, setting $\lambda = 1$ yields a phase boundary between a phase with  small avalanches (percolative phase, $\lambda < 1$), and a phase with large avalanches (bolt phase, $\lambda \ge 1$).
\par
Figure~\ref{fig:PT} shows the phase boundary $V = 7.26 (1+h)/(1-h)$, where the prefactor 7.26 is obtained by fitting the simulation data.
It is difficult to notice the logarithmic drift in the phase boundary due to limited simulation size, however, the mean-field analysis 
suggests that the phase boundary is at $V = 0$ in the limit of $L\to \infty$. Even though the voltage per bond, $V$, goes 
to zero, the externally applied voltage diverges as $L/\log L$. This is analogous to fracture where the stress at failure 
goes to zero, and yet the net applied force at failure diverges in the limit of large length scales \cite{alava2006}. The mean-field theory 
yields a avalanche size exponent of $\tau = 3/2$, which is different from the numerically observed value (Fig.~\ref{fig:LC} a), possibly due to 
the effect of fluctuations. Finally, we have checked that the mean-field theory can also be collapsed by using scaling forms
consistent with the scaling analysis discussed previously (Fig.~\ref{fig:LC} b).
\par
Even though we believe that the phase diagram shown in Fig.~\ref{fig:PT} is qualitatively accurate, there are other possible scenarios that 
cannot be entirely ruled out. It is possible that $V$ is finite at the transition, as suggested by 
the scaling analysis.  
It is also possible that this is an avoided critical point~\cite{footnote6}, {\it i.e.}~large avalanches 
reflecting a crossover to the critical point at $h\rightarrow1,V\rightarrow\infty$.
However, the avalanche size distribution displays a scaling collapse (cf. Fig. 2) and a power law  in a large range. 
Also, the behavior is fairly independent of $h$ for $0.5 < h < 1$, rendering a crossover unlikely.

Our minimal model can be verified experimentally in the following ways: a) 
For high voltages bolt-like avalanches should appear, leading to bolt-like conducting clusters. 
This property has already been observed in Ref.~\cite{chang07}, 
where elongated clusters appear in the presence of finite gate voltage, whereas such anisotropy
is absent when $V=0$~\cite{qazilbash07}. ~b) At 
low contrast ($h \gtrsim 1/2$),
mean resistance jumps and sizes (measured, e.g., using
multiple ESM images) should diverge only at a critical voltage, with
power law distributions $\tau = \alpha\simeq1$. An approach to this regime 
should be easier in hydrostatic pressure-controlled systems like organic materials in the $\kappa$-ET family.

In conclusion, we presented a novel model of avalanches for the metal-insulator transition in VO$_2$,  bringing together recent experimental findings, and also making concrete 
experimental predictions as the relevant parameters are altered.
We have identified a novel continuous transition
controlled by long-range interactions which could be observed in 
particular classes of materials that have evidently smaller contrast, like organic materials under hydrostatic pressure~\cite{kagawa05, singleton02} or bulk V$_2$O$_3$~\cite{limelette03}. Another possibility for achieving low contrast is by tuning hydrostatic pressure, approaching the metal-insulator Ising critical point~\cite{papanikolaou08}.


\begin{acknowledgments}
We would like to thank K.~Dahmen, G.~Durin and S.~Ganapathy 
for enlightening discussions.  S.Z. acknowledges financial support from the short-term 
mobility program of CNR, and A.S., S.P., and J.P.S.~were supported by DOE-BES
DE-FG02-07ER46393. This research was supported in part by the National Science Foundation through TeraGrid resources provided by the
Louisiana Optical Network Initiative (LONI) under grant number TG-DMR100025.
\end{acknowledgments}


\begin{thebibliography}{28}
\bibitem[{foo({\natexlab{a}})}]{footnote3}
\bibinfo{howpublished}{Given the observed metallic domain sizes of 20 nm and
  the sample thickness of $90$~nm, it is clear that avalanches involving
  domains at distances larger than $5$ lattice spacings have a 2D character,
  thereby justifying the use of our 2D model.}
\bibitem[{foo({\natexlab{b}})}]{footnote2}
\bibinfo{howpublished}{The Frank-Condon overlap integral, accounting for the
  $\sim$1\% lattice distortion, would drastically suppress any quantum overlap
  between electronic initial and final states on these length scales ($\sim$20
  nm).}
\bibitem[{foo({\natexlab{a}})}]{footnote5}
\bibinfo{howpublished}{The unit of resistance is $\Delta R=(1-h)R_I$,  for a single resistor. The quantity $R_I$ is defined by the resistance deep in the insulating phase (when all grains are insulating), and here we set it to unity. The unit of voltage and temperature in our simulation is the variance of the random pinning field disorder (assuming $b=1$).  }
\end{thebibliography}
\end{document}